\documentclass[aps,groupedaddress,superscriptaddress,prd, twocolumn]{revtex4}
\usepackage{hyperref}
\usepackage{amsmath,amssymb,amsbsy,amsfonts}
\usepackage{graphicx}
\usepackage{epstopdf}
\usepackage{bm}
\usepackage{soul}
\usepackage{array}
\usepackage{latexsym}
\usepackage{multirow}
\usepackage{sansmath}
\usepackage{subcaption}
\usepackage{lipsum}
\usepackage{float}
\usepackage{xcolor}
\usepackage{comment}
\usepackage{csquotes}	
\usepackage[plain]{algorithm}
\usepackage{algpseudocode}

\newcommand{\mustafa}[1]{{\color{blue}{#1}}}

\renewcommand{\theequation}{\arabic{section}.\arabic{equation}}

\def\be{\begin{equation}}
\def\ee{\end{equation}}
\def\bea {\begin{eqnarray}}
\def\eea {\end{eqnarray}}

\makeatletter
\newcommand\makebig[2]{%
  \@xp\newcommand\@xp*\csname#1\endcsname{\bBigg@{#2}}%
  \@xp\newcommand\@xp*\csname#1l\endcsname{\@xp\mathopen\csname#1\endcsname}%
  \@xp\newcommand\@xp*\csname#1r\endcsname{\@xp\mathclose\csname#1\endcsname}%
}
\makeatother
\makebig{biggg} {3.0}
\makebig{Biggg} {3.5}
\makebig{bigggg}{4.0}
\makebig{Bigggg}{4.5}

\begin{document}

\title{Particle production in a bouncing universe}


\author{Mustafa Saeed} \email{mustafa.saeed@centre.edu}
\affiliation{Centre College,
Danville, KY, USA 40422}

\author{Aiman Nauman} \email{25100114@lums.edu.pk}
\affiliation{School of Science and Engineering, Lahore University of Management Sciences, Lahore 54792, Pakistan}

\author{Irfan Javed} \email{i.javed@unb.ca}
\affiliation{Department of Mathematics and Statistics, University of New Brunswick, Fredericton, NB, Canada E3B 5A3}

\begin{abstract}
\vskip 0.2cm

In massless scalar field cosmology, imposing the universe's physical volume as fundamentally discrete resolves the big bang singularity via a big bounce. We use quantum field theory on curved background to numerically track the number of particles created in the vacuum of a quantum field that propagates through the cosmological bounce. We find that due to geometry's evolution, particle production in all modes initially rises, sharply peaks at the bounce, and varies  slowly afterwards. Further, by comparing with the case of a quantum field propagating on an expanding universe, we discover that the bouncing universe's imprints on quantum matter are distinct: notably, the late time particle production across modes resembles a thermal spectrum. We then use semiclassical gravity and find a similar qualitative result. Here, we also determine how particle production affects geometry's evolution. Our study adds to existing literature on gravity-matter interactions in the context of a bouncing universe, contributes to searches of a bouncing universe's signatures, and suggests---within this context---a link between gravity, quantum fields and thermodynamics.



\end{abstract}
\maketitle
\setlength{\arrayrulewidth}{0.1mm}
\renewcommand{\arraystretch}{2}
\renewcommand{\theequation}{\arabic{equation}}

\hspace{\parindent}


\section{Introduction}

General relativity (GR)---a theory of classical spacetime, matter, and their interactions---enables investigations of black holes and the expanding universe \cite{Wald}. However, GR is not a microscopic theory and is inapplicable at singularities, such as the end point of gravitational collapse of sufficiently massive matter configurations and the cosmological big bang. It is expected that these singularities will be resolved by quantum effects.

These depend on the quantization as different schemes are inequivalent \cite{Halvo}. For example, Schr\"{o}dinger and polymer quantizations use for their configuration spaces the $\mathbb{R}^{n}$ and a lattice, respectively. The two schemes lead to different results when applied to simple systems, such as the harmonic oscillator \cite{Cor07} for example.

The study of quantum matter on a fixed curved geometry is called quantum field theory on curved background (QFTCB) \cite{BirDav82, ParTom09}. A notable result due to QFTCB is that the expansion of the universe creates energy excitations---called particles---in the field vacuum \cite{ParTom09}, a finding which is used to probe physics beyond the standard model, dark matter, and cosmological perturbations \cite{Ko25}. Another finding due to QFTCB is that a black hole radiates like a hot blackbody \cite{Haw75}; it is a link between gravity, quantum fields and thermodynamics. Further, similar to how thermal energy is related to the motion of microscopic particles, a black hole's temperature hints at an underlying microscopic structure.



The formalism that describes the consistent joint propagation of classical geometry and quantum matter is called semiclassical gravity (SG) \cite{Hu20}; its difference from QFTCB lies in that quantum matter now backreacts on the curved geometry, meaning that the latter is no longer fixed as it is in QFTCB. In one version of canonical SG, the gravity phase space variables and matter quantum state are the relevant degrees of freedom that evolve via a state-dependent Hamiltonian constraint and the time-dependent Schr\"{o}dinger equation, respectively \cite{HusSin09,HusSin2,HusMM24}. Application of this SG approach to cosmology has demonstrated how particle production modifies the evolution of classical geometry \cite{HusSin2}.

Quantum geometry, quantum matter, and their interactions are described by quantum gravity (QG) \cite{Ish95, Kie04, Car15}. The resolution of singularities and explanation of gravity's thermodynamic properties are some of QG's goals. There are many QG approaches. As an example, we have got nonperturbative canonical QG described via connection-triad variables. Known as loop quantum gravity (LQG), it mirrors the quantization of the electromagnetic, weak, and strong interactions \cite{AshLew, Ash21}.


LQG motivates loop quantum cosmology (LQC), in which homogeneity and isotropy are imposed on classical gravity expressed in connectrion-triad variables before quantization through a scheme that incorporates spatial discreteness, namely polymer quantization. A notable finding of LQC is the resolution the big bang singularity by replacing it with the big bounce \cite{AshPawSi, AshSi}. The big bounce is also found in effective LQC, which is classical cosmology expressed in connection-triad variables with LQC corrections \cite{Tav, RovEw}.


However, this result is not exclusive to effective LQC variables. Classical cosmology expressed in metric variables including polymer quantum effects also replaces the big bang singularity with the big bounce \cite{Gio22, SaeHus24_2}. This indicates that the big bang singularity resolution follows not from a choice of variables but from a quantization scheme that incorporates discreteness. Bouncing cosmologies are also arrived at in other formalisms, such as higher-order gravity theories, string gas cosmology, and so on; see \cite{NovPer08} for a review.


It is important to search for the signatures of a bouncing universe just as we look for the signatures of the big bang---in the cosmic microwave background (CMB) for example. Work in this direction mostly includes cosmological perturbation theory, in which inflation is preceded by a bounce and a phase of contraction instead of the big bang. Here, the evolution of perturbations across the additional contracting and bouncing phases leads to findings different from those of standard inflation, and they could potentially be verified via the future CMB data \cite{Mie10,Gra10,Lil11,Agu21}. A different work proposes dark matter production through the bounce as a means to detect the the bouncing universe's signatures \cite{Li14}. Another uses the Unruh-DeWitt detector to compare the total particles produced in the vacuum of a conformally coupled massless scalar field in the expanding universe due to GR and the post-bounce universe due to effective LQC \cite{Gar14}.

With a similar motivation of studying gravity-matter interactions in the context of a bouncing universe and identifying its signatures, we numerically study particle production in the vacuum of a massless scalar field that propagates through the bounce using QFTCB and SG. Our setup for both---QFTCB and SG---is inspired by \cite{HusSin2, Mah071, Mah072}, and our work is different from the existing literature on the subject in several aspects. First, we study particle production in a scalar field that propagates through the universe's contracting, bouncing, and expanding phases, not just the post-bounce expansion phase, and secondly, we look at particle production in the vacua of individual field modes. Notably, these extend the work in \cite{Gar14}. Lastly, we generalize to SG to explore how the results differ and also to determine the effect of particle production on the bouncing universe's evolution; this extends \cite{HusSin2}, and is a  direction that does not exist in the literature to the best of our knowledge. 

Using QFTCB, we find that particle production in all modes initially rises, sharply peaks at the bounce, and varies slowly afterwards. At late times, we compare particles produced across modes with corresponding results for a scalar field that propagates on an expanding universe. In the case of an expanding universe, we find that the most particles are produced in the lightest mode. This is aligned with the intuition that lighter modes are easier to excite. For the bouncing universe, given that the field evolves for an equal time pre- and post-bounce, we find that the late-time particle production across modes resembles a thermal spectrum with the most particles produced in an intermediate mode.
Upon generalizing the formalism to SG, we find similar features for particle production and also demonstrate how the quantum matter modifies the  geometry's evolution. These results contribute to searches of the bouncing universe's signatures; suggest a  link between gravity, quantum fields and thermodynamics within the context of a bouncing universe; and provide an example of how quantum matter affects effective geometry.

What follows are a review of effective cosmology in Section \ref{boun_uni}, a study of a quantum scalar field on this effective background using QFTCB and SG in Section \ref{pp_cosm}, and our reflections on the results in Section  \ref{reflns}. 

\section{A bouncing universe}
\label{boun_uni}
In this section, we use \cite{SaeHus24_2} to review cosmological singularity resolution via incorporation of quantum corrections in classical cosmology expressed in the metric variables.    

We look at massless scalar field cosmology with metric 
\begin{align}
\label{eq:metric}
    ds^{2}=-N(t)^{2}dt^{2}+v(t)^{\frac{2}{3}}\left(dx^{2}+dy^{2}+dz^{2}\right),
\end{align}
where $v(t)$ is the cube of the scale factor and $N(t)$ is the lapse. 
The canonical action is
\begin{align}
\label{eq:orac}
S=V_{0}\int{dt \, \left[\left(\frac{1}{16\pi G}\right) \dot{v}p_{v} + \dot{\phi}p_{\phi} - N\mathcal{H}\right]},
\end{align}
where $V_{0}$ is the volume of the cubical fiducial cell,
\begin{align}
V_{0}=\int{d^{3}x};
\end{align}
overdots indicate derivatives with respect to coordinate time; $p_{v}$ and $p_{\phi}$ are the gravity and scalar field momenta, respectively; $\phi$ denotes the scalar field itself;  and $\mathcal{H}$ is the Hamiltonian constraint:
\begin{align}
\label{eq:HamConst}
\mathcal{H}\approx0, \qquad \mathcal{H}=-\left(\frac{1}{16\pi G}\right)\left(\frac{3\,v\,p_{v}{}^{2}}{8}\right)+\frac{p_{\phi}{}^{2}}{2v}.
\end{align} 
While fundamental Poisson brackets,
\begin{align}
\label{eq:PB}
\{v,p_{v}\}=\frac{16\pi G}{V_{0}} \quad \text{and} \quad \{\phi,p_{\phi}\}=\frac{1}{V_{0}},
\end{align}
depend on the fiducial volume, the classical equations of motion do not. These and the Hamiltonian constraint \eqref{eq:HamConst} lead to the Friedmann equation (note that $N$ has been set to $1$ here)
\begin{align}
\label{eq:sffe}
H^{2}=\frac{8\pi G}{3}\rho,
\end{align}
where the Hubble parameter and the field's energy density are
\begin{align}
    H=\frac{\dot{v}}{3v}\quad \text{and}\quad  \rho=\frac{p_{\phi}^{2}}{2v^{2}},
\end{align}
respectively. The Friedmann equation leads to an expanding (contracting) universe, which begins (ends) with a big bang (big crunch); both expanding and contracting cases will be discussed shortly.


This classical theory is expressed in terms of scale-dependent variables. Under the coordinate rescalings
\begin{align}
t\rightarrow t, \qquad x\rightarrow lx,
\end{align}
where $l$ is a dimensionless constant, the parameters and degrees of freedom in the theory rescale as
\begin{align}
\begin{split}
&V_{0} \rightarrow l^{3}\, V_{0},\quad N\ (= 1)\rightarrow N, \quad v \rightarrow l^{-3}\,v,\\
&p_{v} \rightarrow p_{v},\quad\phi \rightarrow \phi, \quad p_{\phi} \rightarrow l^{-3}\, p_{\phi}.
\end{split}
\end{align}
Due to their applicability across scales, we use the above relations to construct scale-invariant variables---such as $V_{0}v$---to express the theory with.


Additionally, to simplify future simulations, we express the theory in terms of dimensionless coordinates, parameters, and variables. In natural units, $\hbar=c=1$, the action is dimensionless, and the terms appearing in the theory have the following dimensions of length $L$:
\begin{align}
\begin{split}
&x=t=L,\quad V_{0}=L^{3}, \quad N=v=1,\\
&p_{v}=\phi=L^{-1},\quad p_{\phi}=L^{-2}.
\end{split}
\end{align}
To make the theory dimensionless, all lengths must be expressed as dimensionless multiples of the Planck length $l_{P}$ using Newton's constant via
\begin{align}
\sqrt{G}=l_{P}.
\end{align}
For example, $G^{-3/2}\,V_{0}\,v$ is dimensionless.

Using the scaling relations and appropriate dimensional assignments, one can define scale-invariant dimensionless quantities to express the theory with. The dimensionless time is $\bar{t}=G^{-1/2}\,t$, and the new canonical degrees of freedom are
\begin{align}
\begin{split}
\bar{v}=&\left( \frac{V_{0}}{16\pi}\right)\left(\frac{1}{\sqrt{G}}\right)^{3}\,v, \qquad
p_{\bar{v}}=\sqrt{G}\,p_{v}, \\
\bar{\phi}=&\sqrt{16\pi G}\, \phi, \qquad p_{\bar{\phi}}=\left( \frac{V_{0}}{\sqrt{16\pi G}}\right)\,p_{\phi}.
\end{split}
\end{align}
Factors of $16\pi$ are included in these definitions for simplicity. We are thus led to the simplified action
\begin{align}
\label{eq:CanAc}
S=\int{d\bar{t} \, \left( \frac{d\bar{v}}{d\bar{t}}\,p_{\bar{v}} + \frac{d\bar{\phi}}{d\bar{t}}\,p_{\bar{\phi}} - \bar{\mathcal{H}}\right)},
\end{align}
where the Hamiltonian is
\begin{align}
\label{eq:HamConstSD}
\bar{\mathcal{H}}=-\frac{3\,\bar{v}\,p_{\bar{v}}{}^{2}}{8}+\frac{p_{\bar{\phi}}{}^{2}}{2\bar{v}}. 
\end{align} 
Hereon, the overbars will be dropped and overdots will indicate time derivatives with respect to dimensionless time. The Hamiltonian \eqref{eq:HamConstSD} and the equations of motion lead to the Friedmann equation
\begin{align}
\label{eq:FRCoe}
H^{2}=\frac{\rho}{6}.
\end{align}
The solution is
\begin{align}
\label{eq:excon}
    v=\frac{\sqrt{27t^{2}}}{6}\,\tilde{p}_{\phi},
\end{align}
where $\tilde{p}_{\phi}$ is a constant of motion. The solution represents an expanding (contracting) universe defined for $t\in \mathbb{R}^{+(-)}$ with $t=0$ representing the big bang (big crunch). 

To arrive at the bounce, it is sufficient to include quantum discreteness corrections to the universe's volume. This can be achieved by polymer quantizing gravity variables via a volume lattice with discretization $\lambda$. We then pick a semiclassical state peaked at the phase space variables to calculate the expectation value of the Hamiltonian operator. Taking the limit that the state is sharply peaked, we get the effective Hamiltonian
\begin{align}
\label{eq:EffHam}
\mathcal{H}_{e}=-\left(\frac{3}{8}\right)\,v\,\left[\frac{\sin(p_{v} \,\lambda/2)}{\lambda/2}\right]^{2}+\frac{p_{\phi}\,^{2}}{2v},
\end{align}
which leads to the effective Friedmann equation 
\begin{align}
\label{eq:EffFri}
    H^{2}=\frac{\rho}{6}\,\left(1-\frac{\rho}{\rho_{c}}\right), \quad \rho_{c}=\frac{3}{8\lambda^{2}}.
\end{align}
This indicates that the universe bounces when $\rho=\rho_{c}$. The solution is
\begin{align}
\label{eq:BouUni}
    v=\frac{\sqrt{27t^{2}+48\lambda^{2}}}{6}\,\tilde{p}_{\phi},
\end{align}
where $\tilde{p}_{\phi}$ is a constant of motion. 


\section{Particle production in cosmology}
\label{pp_cosm}
In this section, we study particle production in a bouncing universe via QFTCB and SG. 
As a preliminary, we review a method to study particle production in a quantum mechanical oscillator with time-dependent mass and frequency using \cite{Mah071, Mah072}. This offers a smooth and useful transition to the study of quantum fields on a cosmological background where we will find that each mode resembles such an oscillator. Then, using QFTCB, we study particle production in the vacuum of a massless inhomogeneous scalar field that propagates on a cosmological background. We determine if a bouncing universe's imprint on quantum matter is distinct by comparing particle production in contracting, expanding, and bouncing universes. Lastly, we use SG to jointly evolve geometry and quantum matter and study how the results differ. Our QFTCB and SG setups are similar to the ones in \cite{HusSin2}.

\subsection{Quantum mechanical time-dependent oscillator}

The Hamiltonian of an oscillator with time-dependent mass and frequency is
\begin{align}
\label{eq:oscham}
    h(t)=\frac{1}{2}\left[\frac{p(t)^2}{m(t)}+m(t)\omega(t)^2q(t)^2\right],
\end{align}
where $(q,p)$ are the oscillator phase space variables and $m$ and $\omega$ are the time-dependent mass and frequency, respectively. Quantizing the oscillator in the position basis leads to the time-dependent Schr\"{o}dinger equation
\begin{equation}
\begin{split}
\label{eq:tdse_tdo}
    \iota\frac{\partial \psi(q,t)}{\partial t}=&\,-\frac{1}{2m(t)}\frac{\partial^2\psi(q,t)}{\partial q^2}\\&\,+\frac{1}{2}m(t)\omega(t)^2\psi(q,t).
\end{split}
\end{equation}
We are interested in a solution of the equation above that can be interpreted as the oscillator's ground state at some $t=t_0$. Therefore, we propose the Gaussian ansatz
\begin{align}
\label{eq:GauAnsO}
    \psi(t, q)=N(t)\exp\left[-R(t)q^2\right],
\end{align}
where $R=m\omega$. The normalization condition gives
\begin{align}
    \label{eq:beta_normalization_osc}   |N|^2&=\sqrt{\frac{2\text{Re}(R)}{\pi}}.
\end{align}
Substituting this ansatz in \eqref{eq:tdse_tdo} gives
\begin{align}
\begin{split}
\label{eq:tdse2}
    \iota\dot{R}=&\,\frac{1}{2m}\left(4R^2-m^2\omega^2\right)\\
    \iota\frac{\dot{N}}{N}=&\,\frac{R}{m}.
\end{split}
\end{align}
For later use, we solve the ODE for $N$ to find
\begin{align}
    N(t)=&\,N_0\exp\left[-\frac{\iota}{2}\int^{t}_{t_{0}}\omega(t')dt'\right]
\end{align}
and also define the excitation parameter:
\begin{align}
    z=\frac{m\omega-2R}{m\omega+2R}.
\end{align}

Due to the time-dependence of the oscillator parameters, we choose the instantaneous energy eigenstates to span the system's Hilbert space at a given time. 
Additionally, for simplicity, we impose that the instantaneous eigenstates at an initial time $t_0$ evolve adiabatically to become the instantaneous eingenstates at a later time $t$. We will refer to this as the adiabatic condition. The set of these eigenstates is given by
\begin{widetext}
\begin{align}
    \psi^n(q,t)=\left(\frac{m\omega}{\pi}\right)^{\frac{1}{4}}\frac{1}{\sqrt{2^nn!}}\,H_n(\sqrt{m\omega}\,q)\exp\left[-\frac{m\omega q^2}{2}-\iota\int^t_{t_{0}}\left(n+\frac{1}{2}\right)w(t')dt'\right],
\end{align}    
\end{widetext}
where $H_n$ are the Hermite polynomials; $n=0,1,2,\ldots$; and the phase is chosen to ensure consistency with the phase of the Gaussian ansatz \eqref{eq:GauAnsO}.

A physically motivated definition for the average number of particles---i.e., energy excitations---in the state at a later time is
\begin{align}
\label{eq:avgp}
    \langle\hat{n}(T)\rangle=\sum_{n=0}^\infty\,n\,|\langle\psi^{n}(T)|\psi(T)\rangle|^{2},
\end{align}
which evaluates to
\begin{align}
    \langle \hat{n} \rangle = \frac{|z|^{2}}{1-|z|^{2}}.
    \label{eq:z_eqn}
\end{align}
There are two other physically motivated justifications for treating $\langle \hat{n} \rangle$ as particle content. Firstly, the average energy evaluates to
\begin{align}
    E(t)=&\,\langle \hat{H}\rangle\\
    =&\,\left(\langle\hat{n}\rangle+\frac{1}{2}\right)\omega(t).
\end{align}
Secondly, \eqref{eq:z_eqn} can be expressed in terms of the Bogoluibov coefficients $\alpha$ and $\beta$, and gives the correct corresponding expression for $\langle \hat{n}\rangle$, as we shall now demonstrate. As a prelude, note that because we are working with the oscillator---which may be interpreted as a field with just one mode---$\alpha$ and $\beta$ will not have mode labels. The treatment first expresses $z$ in terms of $\alpha$ and $\beta$ as
\begin{align}
    z=\frac{\beta^*}{\alpha}e^{-2\iota\dot{\omega}}.
\end{align}
Substituting the equation above in \eqref{eq:z_eqn} gives \begin{align}
    \langle\hat{n}\rangle=\frac{|\beta|^2}{|\alpha|^2-|\beta|^2}.
\end{align}
Since we are working with the oscillator, $|\alpha|^2-|\beta|^2=1$, and this leads to
\begin{align}
    \langle\hat{n}\rangle=|\beta|^2.
\end{align}
This is the expression that comes from using Bogoliubov transformations to compute  $\langle\hat{n}\rangle$ for a quantum field with one mode, which the oscillator resembles. 

We now summarize the takeaways from this subsection. We considered an oscillator with time-dependent mass and frequency. The Hamiltonian of this oscillator is time-dependent. Due to this, the instantaneous eingestates are time-dependent. Further, if one prepared the oscillator in the instantaneous vacuum state at $t_0$, then at a later time, one would find particles in that state. The goal of this section was to review a physically  motivated formalism to study particle production in this context since it resembles that of a quantum field on a cosmological background as we shall now see.



\subsection{Quantum field on a cosmological background}

\subsubsection{Formalism}

An inhomogeneous massless scalar field on an FLRW background described by the metric \eqref{eq:metric} with $N=1$ has the following Hamiltonian in spatial Fourier space:
\begin{equation}
\label{eq:hamphi}
\begin{split}
    \mathcal{H}_{\phi}=&\,\int \frac{d^{3}k}{\left(2\pi\right)^{3}}\,\left(\frac{1}{2}\right)\left[\frac{p_{k}^{2}}{v}+v\left(\frac{k}{v^{\frac{1}{3}}}\right)\phi_{k}^{2}\right]\\
    \equiv &\,\int \frac{d^{3}k}{\left(2\pi\right)^{3}}\,h_{k}.
\end{split}
\end{equation}
Here, $k=|\vec{k}|$ and ($\phi_{k},p_{k}$) are the spatial Fourier transforms of the scalar field phase space variables. Notably, the field modes decouple and $h_k$ resembles $h$ in \eqref{eq:oscham} provided $m=v$  and $\omega=k\,v^{-\frac{1}{3}}$. This demonstrates that each mode 
resembles an oscillator with time-dependent mass and frequency. 
Thus, each field mode is quantized using the prescription from the previous subsection.


Recall that our goal is to study particle production in the vacuum of a field mode as it propagates on a cosmological background. In quantum field theory on curved spacetime, a preferred vacuum exists only when the background admits a global timelike Killing vector, which defines an unambiguous notion of positive-frequency modes. Since generic cosmological spacetimes lack such a symmetry, the concept of vacuum is not unique. The predicted particle content therefore depends on the choice of vacuum, making its selection a crucial ingredient of the analysis.

Our selection is motivated by two perspectives. The first perspective considers that in our canonical approach, the system's Hamiltonian  resembles that of decoupled oscillators. This perspective is agnostic to the additional context that the system represents a quantum field on a cosmological background. Consequently, to study particle production in this system, all steps presented in the previous subsection---including vacuum selection---can be used. 

The second perspective considers the additional context  that this system represents a field on a cosmological background, which lacks a global timelike Killing vector, due to which an unambiguous vacuum selection for the field modes is not possible at $t_0$. This problem can be circumvented if the cosmological background being studied is preceded smoothly by regions that the geometry is flat in. Further, since we are tracking particles produced in a field mode's vacuum, we will also attach a flat piece at the end of each cosmological background, ensuring a clear vacuum choice at the end of our analysis. We note that this engineered (total) background is not a solution to Einstein's equations. However, it enables unambiguous vacuum selection.




In both approaches, the state for each mode is chosen to be the Gaussian
\begin{align}
\label{eq:GauAns}
    \psi_{k}\left(t, \phi_{k}\right)=N_k(t)\exp\left[-R_k(t)\phi_{k}^{2}\right],
\end{align}
where $R_k(t)$ and $N_k(t)$ are time-dependent complex numbers. The normalization condition gives
\begin{align}
    \label{eq:beta_normalization}   |N_k|^2&=\sqrt{\frac{2\text{Re}(R_k)}{\pi}}.
\end{align}
For evolution, we substitute this Gaussian ansatz in the time-dependent Schr\"{o}dinger equation
\begin{align}
i\dot{\psi}_{k}=\hat{h}_{k}\psi_{k},
\end{align}
which leads to
\begin{equation}
\label{eq:TDSE}
\begin{split}
    i\dot{R}_k&=\frac{1}{2v}\left(4R_k^2-v^{\frac{4}{3}}k^2\right),\\
    i\frac{\dot{N_k} }{N_k}&=\frac{R_k}{v}.
\end{split}
\end{equation}

We initialize these modes at $t=t_0$ in their respective Hamiltonian's instantaneous ground state. 
To set these initial conditions, we compare \eqref{eq:GauAns} with the ground state of an oscillator,
\begin{align}
\psi(x)=\left(\frac{mw}{\pi}\right)^\frac{1}{4}\exp\left(-\frac{mw}{2}q^2\right),
\end{align}
and use the correspondences $m=v$ and $\omega=kv^{-\frac{1}{3}}$ to fix
\begin{align}
    R_k(t_0)=\frac{kv(t_0)^{\frac{2}{3}}}{2}, \quad N_k(t_{0})=\left(\frac{kv(t_0)}{\pi}\right)^{\frac{1}{4}}.
\end{align}


To compute particle production at a later time $T\geq t_{0}$, the Hamiltonian's instantaneous eigenstates $\psi^{n}_{k}(t)$ are required. These are computed from the eigenstates at time $t = t_{0}$ via the adiabatic approximation for simplicity. Then, average particle number in $\psi_{k}$ at time $t=T$ is
\begin{align}
    \langle \hat{n}_{k}(T)\rangle =\sum_{n_{k}=0}^{\infty}\,n_{k}|\langle\psi_{k}^{n}(T)|\psi_{k}(T)\rangle|^{2},
\end{align}
which evaluates to
\begin{align}
    \langle \hat{n}_{k} \rangle = \frac{|z_{k}|^{2}}{1-|z_{k}|^{2}}
\end{align}
with 
\begin{align}
    z_{k}= \frac{v^{\frac{2}{3}}k-2R_k}{v^{\frac{2}{3}}k+2R_k};
\end{align}
see \cite{HusSin2, Mah071, Mah072} for details. As expected, at $t=t_0$,
\begin{align}
    \langle \hat{n}_{k} \rangle = 0.
\end{align}
Using this proposal, cosmological particle production in the vacuum of a quantum field can be computed at any arbitrary time. For application of this proposal to studying particle production in an asymptotically flat universe described by $a(t)\sim \tanh(t)$, and a radiation dominated universe with $a(t)=t^{\frac{1}{2}}$, see \cite{HusSin2}.

Next, we will apply this formalism to study particle production in the contracting and expanding universes described by \eqref{eq:excon} in addition to the bouncing universe given by \eqref{eq:BouUni}. This is in line with the first perspective mentioned earlier. Then, we will smoothly attach pieces of flat spacetimes at the start and end of each background to ensure unambiguous initial and final vacua; this is in line with the second perspective. 



\subsubsection{Universes from classical and effective theories}
For the contracting universe, we performed the simulation  from $t=-\tau$ through $t=-2$ 
with $\tau=800$, and a time step of $\Delta t = 0.01$ (the results do not change with lower step sizes). The reason for performing the simulation till $t=-2$ instead of the more natural $t=-1$ now follows. Recall that---in line with the second perspective---we will later smoothly attach pieces of flat spacetimes at the beginning and end of the geometries we study. To ensure smooth transitions between the geometry and flat spacetime, we use sine functions.
In the case of a sine function being attached at the end of the contracting universe, we want to avoid the sine function bringing the volume to or below zero, which does occur if we end this simulation at $t=-1$. This justifies using $t=-2$. We also selected $\tilde{p}_{\phi}=0.9946$ and will justify this choice when we discuss semiclassical gravity. 

\begin{figure*}[htbp!]
\includegraphics[width=\linewidth]{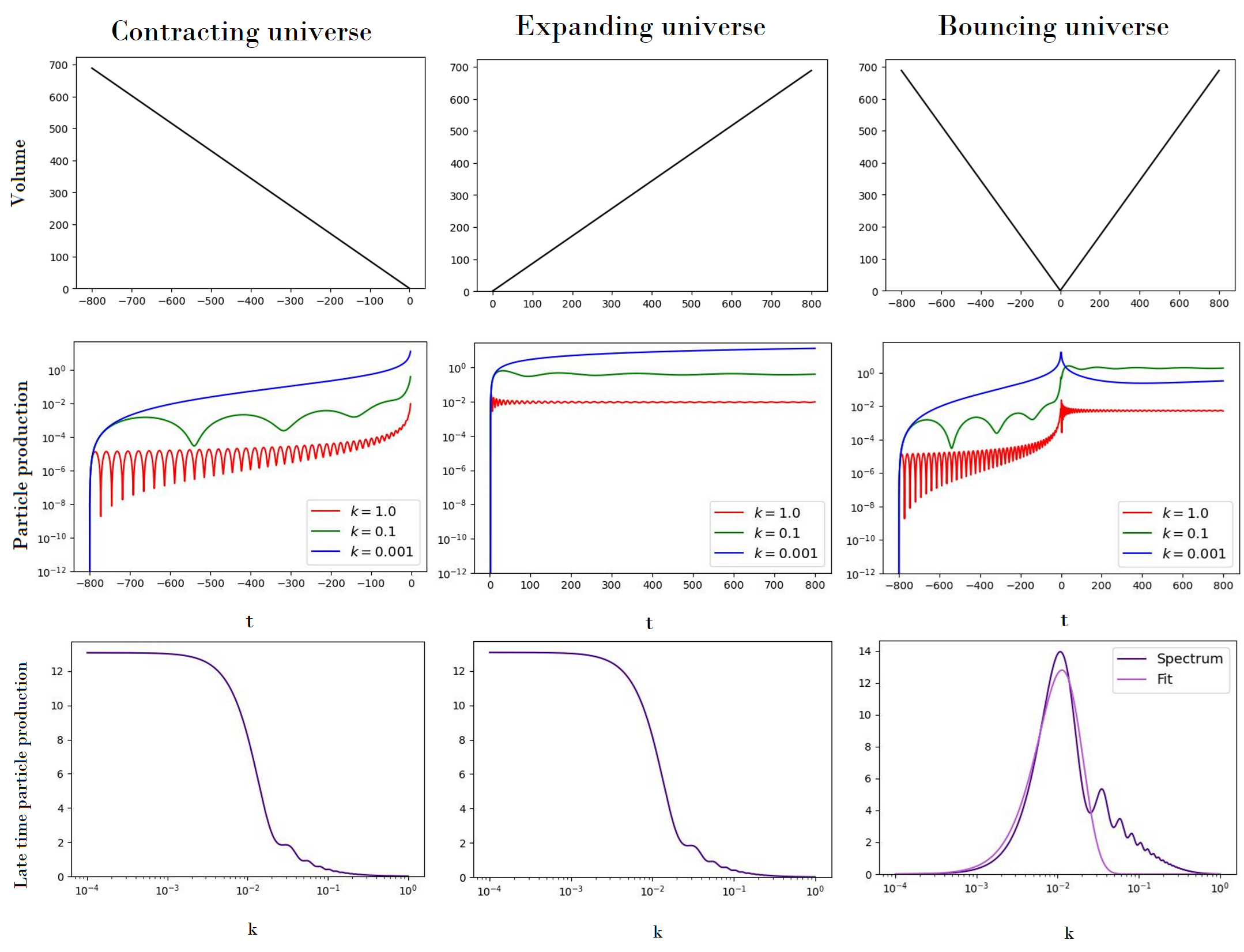}
\caption{\label{fig:figure1} Particle production in the contracting, expanding and bouncing universes}
\end{figure*}

The results are in the left column in figure \ref{fig:figure1}. 
The top subfigure shows the contracting universe, and the central one displays particle production in some modes. We note that particle production occurs in all modes. Specifically, it rises sharply after initialization and when the curvature reaches a maximum. Further, the number of particles at late times is larger for lighter modes. These results are aligned with the intuition that a varying geometry creates particles and that the lighter modes are easier to excite. This trend is more apparent in the bottom subfigure, which is particle production spectrum across $10^{5}$ modes at the final time.

For the expanding universe, we performed the simulation from $t=2$ till $t=\tau$ with time step $\Delta t = 0.01$. The reason for choosing the initial time $t=2$ is like the reason for choosing a final time of $t=-2$ in the case of the contracting universe. 
The results are in the central column in figure \ref{fig:figure1}. As before, the top subfigure is a plot of the geometry, the central one displays particle production in some modes. Here too, particle production occurs in all modes; more precisely, it rises sharply after initialization and varies steadily afterwards. The rise is sharper compared to the previous case because the field experiences high curvature in the beginning. However, the late time particle production spectrum---shown in the bottom subfigure---is identical.

For a bouncing universe, we performed the simulation from $t=-\tau$ till $t=\tau$ with volume discretization $\lambda=1$ and $\Delta t = 0.01$. The results are in the right column of figure \ref{fig:figure1}. 
The top subfigure displays the evolution of volume. The central one shows particle production in the selected modes. It indicates that particle production initially rises, sharply peaks at the bounce, and then varies steadily. Notably, the number of particles for $k=0.001$ is lesser than that for $k=0.1$ despite the former being the lighter mode. This trend is apparent in the bottom subfigure, which shows the particle production spectrum across $10^5$ modes at the final time. It is qualitatively different from those of the contracting and expanding universes. Specifically, the spectrum shows suppression of particle production in the lighter and heavier modes with a peak in the middle. 

Additionally, the spectrum roughly resembles a Planckian blackbody spectrum. Therefore, we try to fit 
a Planck blackbody inspired formula
\begin{align}
    u_{A, T}(k) = \frac{Ak^{3}}{e^{\frac{k}{T}}-1}
\label{planck_form}
\end{align}
to it, where $T$ and $A$ is are the fitting parameters to be determined. $T$ and $A$ may be interpreted as temperature and an overall normalization respectively. 
The result of fitting (\ref{planck_form}) to spectral data gives the fitting parameters $A=1.4\times10^8$ and $T=4.0\times 10^{-3}$ with the coefficient of determination $R^{2}=0.88$; the lighter plot in the bottom subfigure is the fitted curve. Notably, the spectrum's ``temperature" is sub-Planckian. 

\begin{figure*}[htbp!]
\includegraphics[width=0.4\linewidth]{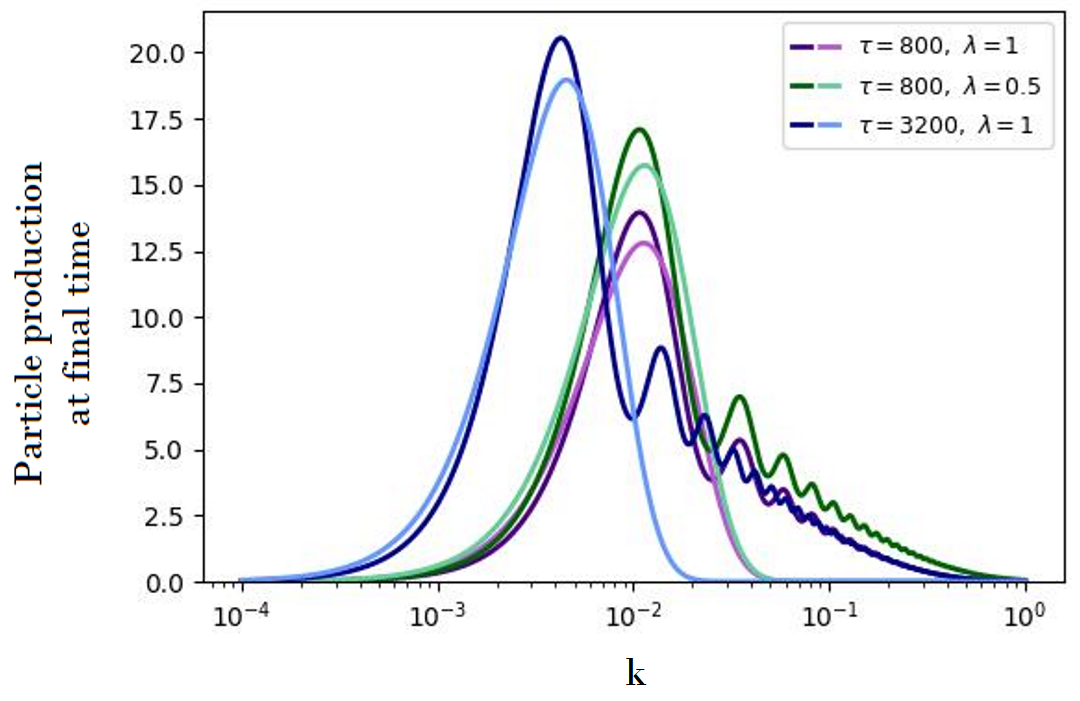}
\caption{\label{fig:figure2}Late time particle production spectrum in bouncing universes with different $\lambda$ and $\tau$}
\end{figure*}

Results of fitting Eq.~(\ref{planck_form}) to spectra with varying $\lambda$ and $\tau$ are shown in figure \ref{fig:figure2}; here, the darker colors are used to plot the spectra themselves, and the lighter ones are used for the fits. The table below lists the corresponding $A,\, T$ in addition to $R^{2}$.
\begin{table}[h]
\centering
\begin{tabular}{| ccccc |}
\hline
$\tau$ & $\lambda$ & $T$ & $A$ & $R^2$ \\
\hline
$\,800\,$  & $\,1\,$ & $\,4.0\times 10^{-3}\,$ & $\,1.4\times10^{9}\,$  & $\,0.88\,$ \\
$\,800\,$  & $\,\displaystyle{\frac{1}{2}}\,$ & $\,4.0\times 10^{-3}\,$ & $\,1.7\times10^{9}\,$  & $\,0.85\,$ \\
$\,3200\,$ & $\,1\,$ & $\,1.6\times 10^{-3}\,$ & $\,3.2\times10^{10}\,$ & $\,0.81\,$ \\
\hline
\end{tabular}
\caption{Results for different values of $\tau$ and $\lambda$.}
\label{tab:resultssg}
\end{table}
Notably, more particles are produced for smaller $\lambda$. This is in line with the intuition that smaller $\lambda$ implies smaller volume and higher curvature at the bounce, which produces more particles. More particles are also produced for larger $\tau$. This is to be expected since the field propagates on the varying geometry for longer. 

A related observation is that as $\tau$ decreases---whereby the contraction and expansion phases increasingly deviate from their classical counterparts---$T$ increases. 
As a further check, we verified that for $\tau=1$, $A=1.9\times 10^{-1}$, $T=3.7\times 10^{-1}$ with $R^2=0.93$. This suggests that for a quantum universe, the spectrum's ``temperature" is nearly Planckian. 


We now state the main takeaway from this analysis: the late time particle production spectrum due to a bouncing universe is qualitatively distinct from that due to  a contracting or expanding universe. Notably, this spectrum resembles that of a blackbody's. Further, we demonstrate how the height of the spectrum's peak and $k_{\text{peak}}$ depend on  $\tau$ and $\lambda$. Recall, that this analysis used the resemblance of a mode's Hamiltonian to that of an oscillator's to initialize all field modes in the vacuum despite the initial geometry's not being flat. Next, we will perform this analysis in the case of similar contracting, expanding, and bouncing universes that are flat in the asymptotic past and future as well. We will note how the results differ qualitatively, if at all.



\subsubsection{Universes with preferred initial and final vacua}

In this section, we alter the studied cosmological backgrounds by smoothly attaching flat pieces at the beginning and end using sine functions.  We note that these backgrounds are not solutions to Einstein's equations. However, they enable an unambiguous selection for the vacuum at the initial and final times of the analysis. With these choices, we can initialize a field mode in the vacuum and note its final particle content at times when we have a preferred vacuum choice.

\begin{figure*}[htbp!]
\includegraphics[width=\linewidth]{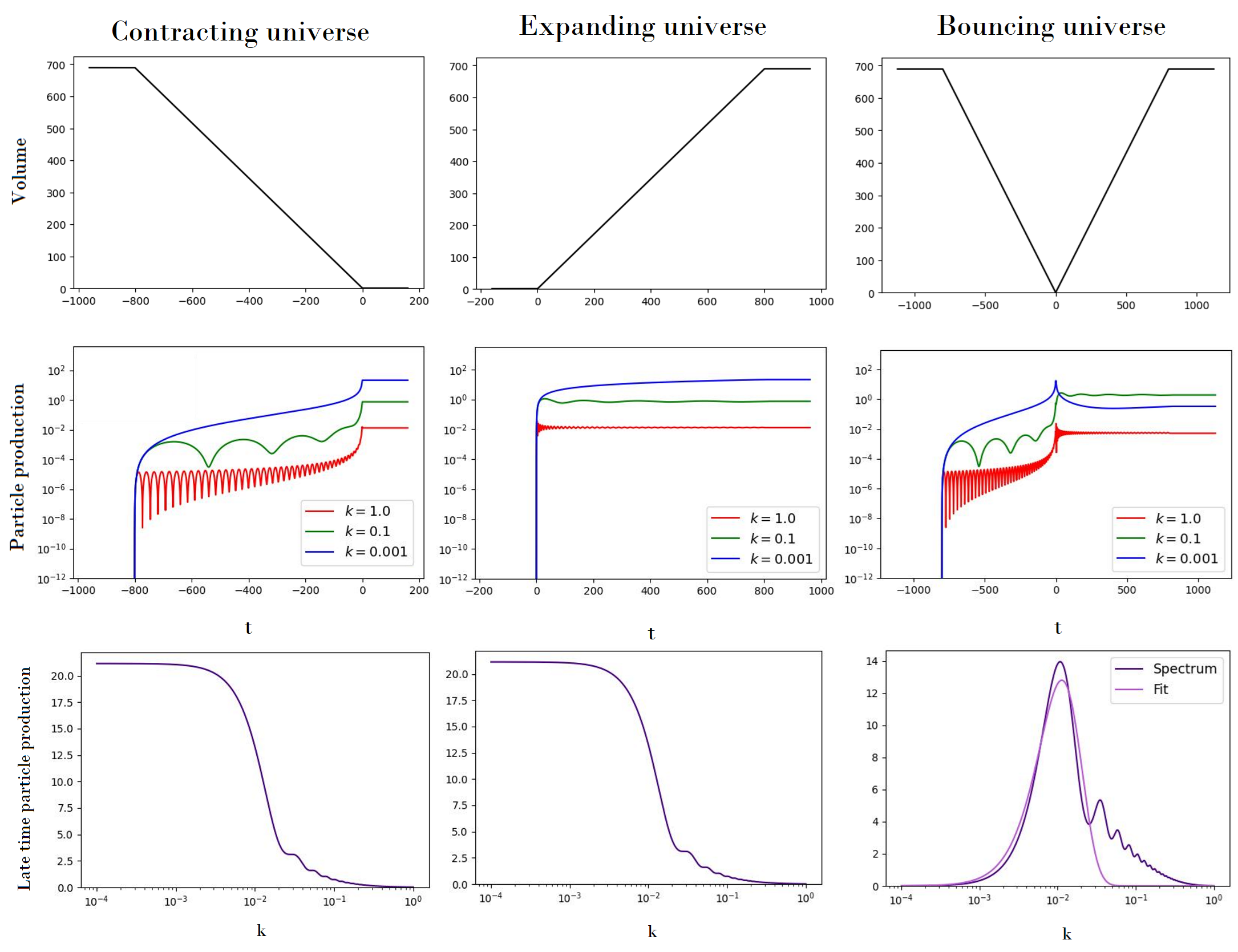}
\caption{\label{fig:figure3}Particle production in the asymptotically flat contracting, expanding and bouncing universes}
\end{figure*}

Our results are in figure \ref{fig:figure3}. The left column has results relevant to the contracting universe (with the abovementioned flat spacetime pieces). The top subfigure shows the volume plot. It starts flat. Then, at $t=-\tau$, where $\tau=800$, it smoothly transitions to the contracting geometry given by \eqref{eq:excon}. The geometry contracts till $t=-2$ after which it smoothly becomes flat again. The middle subfigure shows particle production in some modes. The only way the results differ from the corresponding ones in figure \ref{fig:figure1} is that particle production is constant when the geometry is flat---as expected. Specifically, before the contraction starts, there are no particles produced. Then, once the geometry flattens at the end, particle production stops. The bottom subfigure shows the late time particle production spectrum which has the same qualitative features as seen previously. The overall larger number of particles produced has to do with the field propagating through a volume profile that has a sine piece at roughly Planck volume.

The middle column shows results relevant to the expanding universe with the flat pieces. The top subfigure has the geometry's plot: it is the corresponding geometry from figure \ref{fig:figure1} smoothly preceded and followed by flat pieces. Particle production in select modes that propagate on this geometry show similar qualitative features as before with the only expected difference being that no particles are produced when the geometry is flat. Consequently, the late time particle production spectrum is also similar as the corresponding one from figure \ref{fig:figure1}. Overall, more particles are produced for this volume profile compared to the expanding universe without the flat pieces. This is again because this volume profile has a sinusoidal piece at approximately Planck volume.

The right column has results relevant to the bouncing universe that is flat in the past and future. The top subfigure shows the geometry's plot: it is the bouncing profile from figure \ref{fig:figure1} smoothly attached to flat pieces at both ends. The middle subfigure shows particle production in some modes. Note that it retains the same features as the corresponding plot from figure \ref{fig:figure1} with the (expected) difference being that particles are not produced when the geometry is not varying. Lastly, the late time particle production spectrum retains the characteristic shape from before. The fitting parameters are $A=1.4\times10^8$ and $T=4.0\times 10^{-3}$ with $R^2=0.88$. They are---to these significant figures---identical to those found from figure \ref{fig:figure1}. Further, the overall number of particles produced at late times for this volume profile is similar to that produced in a bouncing universe without the flat pieces. This is because the sinusoidal pieces here are not at an approximate Planck volume; they occur at a much larger volume, which does not spur a lot of particle production owing to low spacetime curvature.

Our main takeaway from this analysis now follows. Smoothly attaching flat pieces at the beginning and end of a cosmological background enables initializing a field mode and noting its final particle content in preferred vacua. Our analysis demonstrates that even with this choice of geometries, the bouncing universe's late time particle production signature is qualitatively distinct from that of a contracting or expanding universe. In what follows thus, we do not append the flat pieces to our backgrounds: doing so does not alter the qualitative features of our results.

\subsubsection{Other bouncing universes}

Using QFTCB, we also study particle production due to other bouncing universes described by an engineered $v(t)$. The selected $v(t)$ are not found by solving Einstein's equations for some matter type. They are useful because they allow us to probe the properties of the late-time spectrum further. The cases we explored are represented by the following.
\begin{widetext}
\begin{align*}
    \text{Single Gaussian bounce I:} \quad v(t)=&\,\bar{v}-(\bar{v}-v_{c})\,\exp\left(-\frac{t^{2}}{2\,\left(\sigma\tau\right)^2}\right),\\
    \text{Single Gaussian bounce II:} \quad v(t)=&\,\bar{v}+(\bar{v}-v_{c})\,\exp\left(-\frac{t^{2}}{2\,\left(\sigma\tau\right)^2}\right),\\
    \text{Single quadratic bounce I:} \quad v(t)=&\,(\bar{v}-v_{c})\left(\frac{t}{\tau}\right)^2 + v_{c},\\
    \text{Single quadratic bounce II:} \quad v(t)=&\,(v_{c}-\bar{v})\left(\frac{t}{\tau}\right)^2+2\bar{v}-v_{c},\\
    \text{Two sinusoidal bounces I:} \quad v(t)=&\,(\bar{v}-v_{c})\sin\left(\frac{\pi t}{\tau}\right)+\bar{v},\\
    \text{Two sinusoidal bounces II:} \quad v(t)=&\,(v_{c}-\bar{v})\sin\left(\frac{\pi t}{\tau}\right) +\bar{v}.\\
\end{align*}
\end{widetext}
Here, the universes evolve from $t=-\tau$ till $t=\tau$, $\bar{v}$ is the initial and final volume for all universes, $v_{c}$ is the critical Planckian volume, and $\sigma$ controls the width of the Gaussian bounces. We choose $\sigma=1/4$ to ensure that $v(t=\pm \tau)\approx \bar{v}$. The universes labeled I initially contract and bounce at critical volume $v_{c}$, after which they expand. While the Gaussian and quadratic universes return to $\bar{v}$ after a single bounce, the sinusoidal one undergoes an additional bounce at $2\bar{v}-v_{c}$ before returning to $\bar{v}$. The universes labeled II initially expand and bounce at large volume $2\bar{v}-v_{c}$, after which they contract. Just as in the case of the universes labeled I, the Gaussian and quadratic universes return to $\bar{v}$ after a single bounce whereas the sinusoidal one does the same after undergoing an additional bounce at $v_{c}$. Notably, all of these selected geometries are neither smoothly preceded nor followed by flat pieces because we established in the previous analysis that this selection does not alter the qualitative results.

\begin{figure*}[htbp!]
\includegraphics[width=\linewidth]{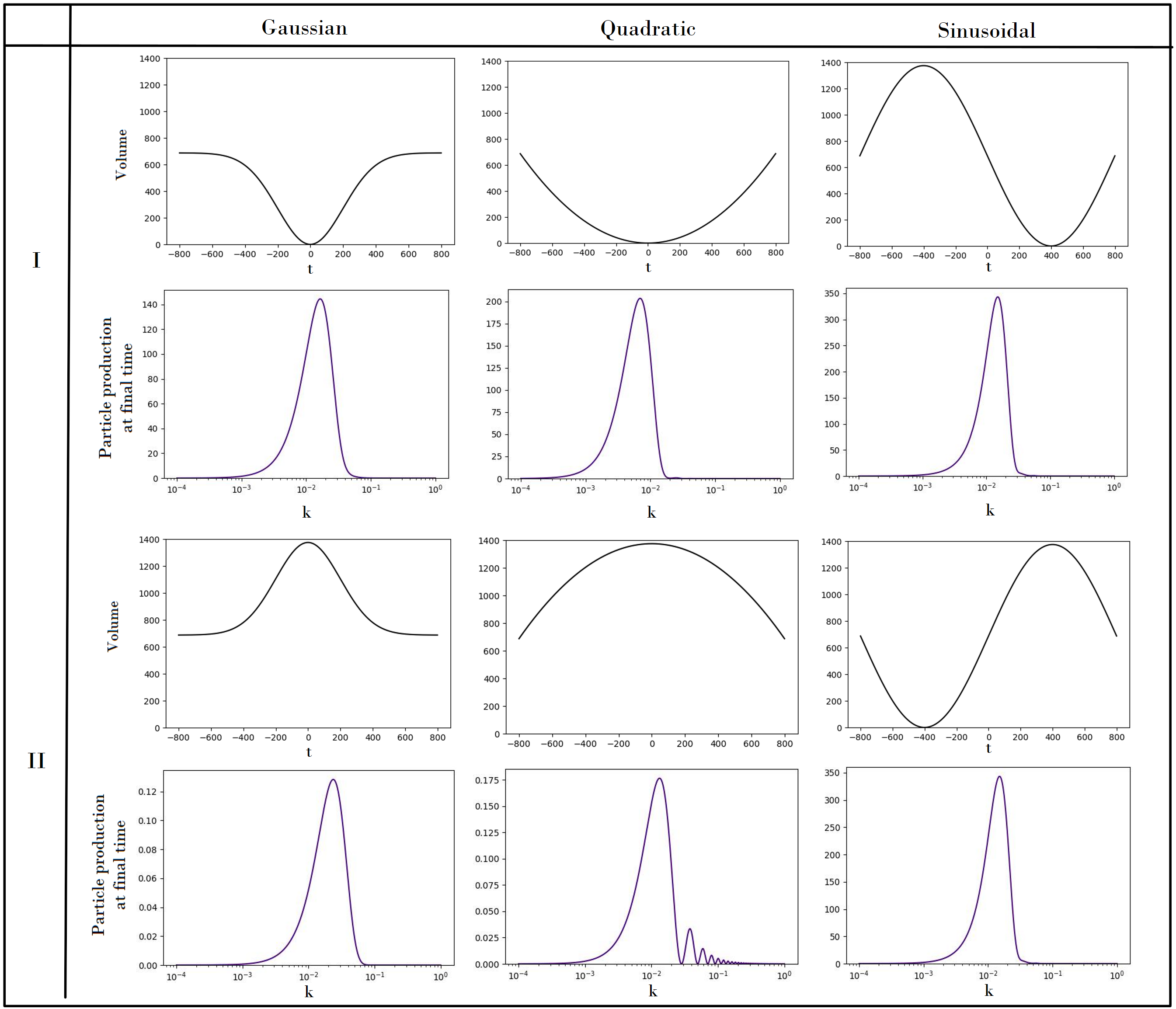}
\caption{\label{fig:figure4}Late time particle production spectrum for different bouncing universes}
\end{figure*}

To compare with the effective universe results from figure \ref{fig:figure1}, we choose the same $\tau$ and $\bar{v}$. The results are in figure \ref{fig:figure4}.
That the characteristic spectrum like the bouncing universe's from figures \ref{fig:figure1} and \ref{fig:figure3} is present in all cases suggests that a condition for such a spectrum to be produced is a bounce---regardless of the volume at which the bounce occurs. Relatedly, the spectrum shape is not due to the field experiencing high curvature. Curvature affects only the peak on the spectrum. This is shown via comparisons of Gaussian I with II and Quadratic I with II: the IIs have much less particle production due to the field experiencing a lower overall curvature. Further verification comes from the case of multiple bounces---compare sinusoidal I with II. These have a high and a low curvature bounce. It appears that the net number of particles produced is similar in both because the field in both cases experiences the same curvature overall no matter whether it sees expansion or contraction first (equivalently, a low or high curvature first).

In the next section, we will generalize our formalism to SG and study how the results differ.


\subsection{Semiclassical cosmology}
\subsubsection{Formalism}
A system of effective cosmological geometry and massless (classical) homogeneous scalar field has Hamiltonian constraint \eqref{eq:EffHam}, which leads to the effective Friedmann equation \eqref{eq:EffFri}. Our semiclassical cosmological model comprises the effective cosmological geometry and a quantum massless inhomogeneous scalar field. The state-dependent effective Hamiltonian constraint for this system is
\begin{align}
\label{eq:ScEffHam}
\mathcal{H}_{e,\psi}=-\left(\frac{3}{8}\right)\,v\,\left[\frac{\sin(p_{v} \,\lambda/2)}{\lambda/2}\right]^{2}+v\,\langle\hat{\rho}\rangle_{\psi},
\end{align}
where 
\begin{align}
    \hat{\rho}=&\,\frac{1}{v}\,\int \frac{d^{3}k}{\left(2\pi\right)^{3}}\,\left(\frac{1}{2}\right)\left[\frac{\hat{p}_{k}^{2}}{v}+v\left(\frac{k}{v^{\frac{1}{3}}}\right)\hat{\phi}_{k}^{2}\right]\\
    \equiv&\,\frac{1}{v}\,\int \frac{d^{3}k}{\left(2\pi\right)^{3}}\,\hat{h}_{k}.
\end{align}
The geometric phase space variables and matter state evolve via Hamilton's equations and the time-dependent Schr\"{o}dinger equation, respectively:
\begin{align}
\dot{v}=\{v,\mathcal{H}_{e,\psi}\}, \quad \dot{p}_{v}=\{p_{v},\mathcal{H}_{e,\psi}\}, \quad i\dot{\psi}_{k}=\hat{h}_{k}\psi_{k}.
\end{align}
These preserve the Hamiltonian constraint in time:
\begin{align}
    \dot{\mathcal{H}}_{e}=0.
\end{align}
The Hamiltonian constraint and the system of equations are equivalent to the effective Friedmann equation
\begin{align}
\label{eq:SDEffFri}
    H^{2}=\frac{\langle\hat{\rho}\rangle_{\psi}}{6}\left(1-\frac{\langle\hat{\rho}\rangle_{\psi}}{\rho_{c}}\right)
\end{align}
and the time-dependent Schr\"{o}dinger equation. Initial data $\{v(t_{0}),\psi_{k}(t_{0})\}$ are used to compute $\langle\hat{\rho}\rangle_{\psi}$. Then, the coupled first-order equations are solved for $v$ and  $\psi_{k}$ at the next time step, and the process is repeated.

For the Gaussian ansatz \eqref{eq:GauAns},
\begin{align}
\label{eq:ModSumEneDen}
    \langle \rho\rangle_{\Psi}&=\frac{1}{v}\int
    \frac{d^3k}{(2\pi)^3}\frac{1}{2\text{Re}(R_k)}\left[\frac{v}{4}\left(\frac{k^2}{v^{\frac{2}{3}}}\right)+\frac{|R_k|^2}{v}\right],
\end{align}
and the time-dependent Schr\"{o}dinger equation leads to \eqref{eq:TDSE}. Therefore, it is sufficient to use \eqref{eq:SDEffFri} with \eqref{eq:ModSumEneDen} and \eqref{eq:TDSE} to evolve $v$ and $R_k$, respectively. This joint evolution of effective geometry and quantum matter captures how geometry produces particles and how particle production affects the geometry.

We also note that $\dot{v}_0$---found via \eqref{eq:SDEffFri}---is dependent on $v_0$ and consequently $\tilde{p}_\phi$. Our selection of $\tilde{p}_\phi=0.9946$ (to four significant figures) satisfies the condition that the volume profiles in QFTCB and SG have the same $v_0$ and $\dot{v}_0$ for the same $t_0$. 
This ensures that any differences in SG volume evolution from QFTCB's are solely due to particle production backreaction.


\subsubsection{Results}
For simulations, we used $\lambda=1$ and replaced the integral in Eq.~(\ref{eq:ModSumEneDen}) by a sum with $10^{5}$ equispaced $k$-values between $0$ and $1$ on the log scale, the $1$ here representing an ultraviolet cutoff. We chose the log scale due to its utility in presenting the spectrum graphs; we also determined that this selection leads to similar results as the one with equispaced $k$ on the $k$ scale itself.

At the starting time $t=-\tau$ with $\tau=800$, the initial data were $\{v_{0},\psi_{k}^{0}\}$, where $v_{0}$ is evaluated via \eqref{eq:BouUni}.
To compare with the QFTCB case, we performed the simulation for equal contraction and expansion times. 

\begin{figure*}[htbp!]
\centering
\includegraphics[width=0.5\linewidth]{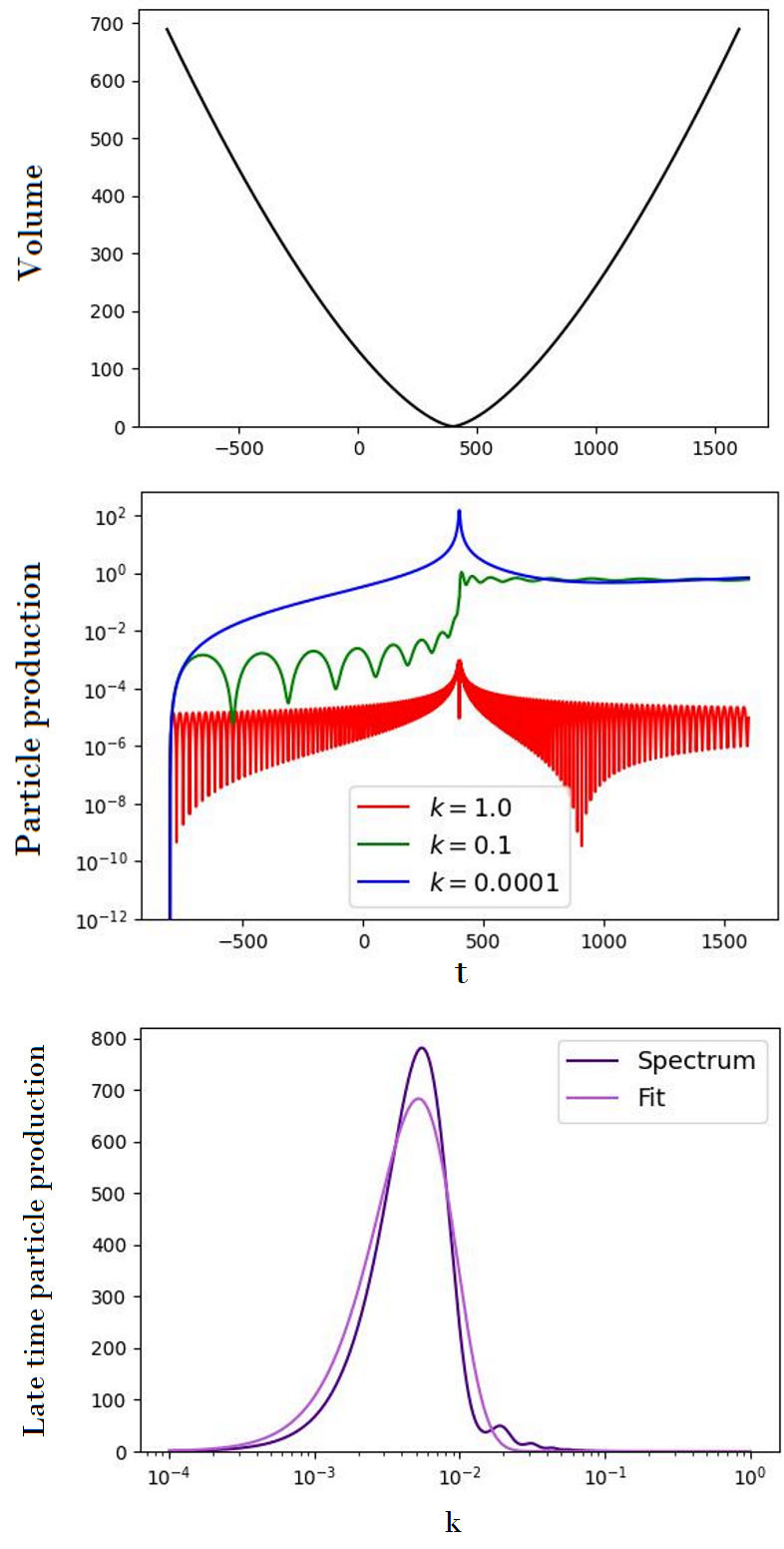}
\caption{\label{fig:figure5}Particle production through a cosmological bounce in SG}
\end{figure*}

The results are in figure \ref{fig:figure5}. 
The joint evolution of quantum matter and effective geometry causes a bounce that occurs later regardless of the same $v_0$ and $\dot{v}_0$ in both (QFTCB and SG) cases.
Further, particle production in the selected modes is qualitatively similar to the QFTCB case. Some differences include more oscillations in time and the larger number of particles produced overall. The late-time particle production spectrum  in the bottom subfigure shows features similar to the one from QFTCB with some differences being no oscillations for the heavier modes and a larger value for overall particle production. Fitting the same formula as before to spectral data yields the parameters $A= 7.7\times10^{10}$ and $T= 1.8\times 10^{-3}$ with $R^2= 0.97$.

These qualitative results are unchanged if one smoothly attaches flat spacetime pieces both to the beginning and to the end of the evolving volume metric variable and evolves the field modes through these additional pieces. Further, the effects of changing $\lambda$ and/or $\tau$ in the spectrum are the same as those demonstrated in figure \ref{fig:figure2} in the QFTCB case: increasing $\lambda$ raises the spectrum's peak, and increasing $\tau$ raises the peak and shifts it toward lighter $k$.

A similar set of SG equations can be used to study classical cosmology provided  \eqref{eq:SDEffFri} resembles the classical---instead of the effective---Friedmann equation. In this case, the late time spectrum is similar to the corresponding one arrived at from QFTCB; for details, see \cite{HusSin2}.

\section{Reflections}
\label{reflns}
A resolution of the big bang singularity comes from including quantum discreteness in the universe's volume. The resultant effective theory replaces the big bang singularity by a big bounce. We studied particle production in a quantum scalar field as it propagates across the cosmological bounce using QFTCB and SG with the goal of identifying the bounce's signatures. 

The proposed formalism for QFTCB used the Schr\"{o}dinger picture, whereby the state for each field mode was evolved by its time-dependent Hamiltonian. 
For SG, we used a state-dependent Hamiltonian constraint and the Schr\"{o}dinger equation for each field mode to jointly evolve the effective geometry and quantum matter, respectively. With this SG proposal, we studied how particle production and effective geometry influence each other.

Our results show that particle production in a bouncing universe is distinct from that in one that only expands or contracts. Moreover, the particle production across modes at late times roughly resembles a thermal spectrum, to which we fit a formula (see Eq.~(\ref{planck_form})) inspired by the Planck blackbody radiation formula and determined the fitting parameters $A$ and $T$. The latter may be seen as a ``temperature," and the goodness of the fit---given by the coefficient of determination $R^{2}$---was used as a comparison between the QFTCB and SG results. It is apparent that the fit was much better with the quantum backreaction of matter incorporated, or in the SG case. With SG, we also found that the geometry's evolution is quite different from the case where there is no matter backreaction. Therefore, this study adds to the existing literature on gravity-matter interactions in the context of a bouncing universe, contributes to searches for the bounce's signatures, and suggests---within this context---a link between gravity, quantum fields and thermodynamics.

Notably, we use the Gaussian ansatz for the matter state because it is simple and enables inexpensive computations. We do emphasize though that this does not represent any approximation whatsoever. The evolution of this ansatz is an exact quantum evolution. 


Future directions of exploration include determining the extent to which $T$ is an actual temperature, identifying the physical mechanism that selects the mode at which the spectrum peaks and investigating how our results may translate into cosmological observations. A more fundamental quantum cosmological model may be required for this analysis. 
Other avenues of exploration include generalizations of the current model by incorporating the cosmological constant, anisotropies, and scalar field mass. Relatedly, one may also use matter with nonzero spin to investigate the cosmological abundance of particles for each spin type.


{\bf Acknowledgements} We thank Viqar Husain, Suprit Singh and Edward Wilson-Ewing for helpful discussions. MS is supported by the FDC award at Centre College. IJ is supported by NSERC.  

\bibliography{ref}

\end{document}